\documentclass[twocolumn,tighten,times]{aastex61}

\newcommand{\src}{G1.9+0.3}
\catcode`\@=11
\newcommand{\gapprox}{\mathrel{\mathpalette\@versim>}}
\newcommand{\lapprox}{\mathrel{\mathpalette\@versim<}}
\newcommand{\propapprox}{\mathrel{\mathpalette\@versim\propto}}
\newcommand{\@versim}[2]
  {\lower3.1truept\vbox{\baselineskip0pt\lineskip0.5truept
\ialign{$\m@th#1\hfil##\hfil$\crcr#2\crcr\sim\crcr}}}
\catcode`\@=12

\defcitealias{borkowski14}{B14}
\defcitealias{carlton11}{C11}
\received{January 11, 2017}
\revised{February 8, 2017}
\accepted{February 16, 2017}
\journalinfo{Astrophysical Journal Letters}

\shorttitle{EXPANSION OF YOUNGEST GALACTIC SNR G1.9+0.3}

\begin{document}

\title{Asymmetric Expansion of the  
Youngest Galactic Supernova Remnant G1.9+0.3}

\correspondingauthor{Kazimierz J. Borkowski}
\email{kborkow@ncsu.edu}

\author{Kazimierz J. Borkowski}
\affiliation{Department of Physics, North Carolina State University, 
Raleigh, NC 27695-8202, USA}

\author{Peter Gwynne}
\affiliation{Department of Physics, North Carolina State University, 
Raleigh, NC 27695-8202, USA}

\author{Stephen P. Reynolds}
\affiliation{Department of Physics, North Carolina State University, 
Raleigh, NC 27695-8202, USA}

\author{David A. Green}
\affiliation{Cavendish Laboratory;
  19 J.J. Thomson Ave., Cambridge CB3 0HE, UK}

\author{Una Hwang}
\affiliation{Department of Astronomy, University of Maryland, 
  College Park, MD 20742, USA}

\author{Robert Petre}
\affiliation{NASA/GSFC, 
  Code 660, Greenbelt, MD 20771, USA}

\author{Rebecca Willett}
\affiliation{Department of Electrical and Computing Engineering, 
  University of Wisconsin-Madison,
  Madison, WI 53706, USA}

\begin{abstract}

The youngest Galactic supernova remnant (SNR) \object{G1.9+0.3},
produced by a (probable) SN Ia that exploded $\sim 1900$ CE, is
strongly asymmetric at radio wavelengths, much brighter in the north,
but bilaterally symmetric in X-rays.  We present the results of X-ray
expansion measurements that illuminate the origin of the radio
asymmetry.  We confirm the mean expansion rate (2011 to 2015) of
0.58\%\ yr$^{-1}$, but large spatial variations are present.  Using
the nonparametric ``Demons" method, we measure the velocity field
throughout the entire SNR, finding that motions vary by a factor of 5,
from $0 \farcs 09$ to $0 \farcs 44$ yr$^{-1}$.  The slowest shocks are
at the outer boundary of the bright northern radio rim, with
velocities $v_s$ as low as 3,600 km s$^{-1}$ (for an assumed distance
of 8.5 kpc), much less than $v_s = 12,000$ -- $13,000$ km s$^{-1}$
along the X-ray-bright major axis.  Such strong deceleration of the
northern blast wave most likely arises from the collision of SN ejecta
with a much denser than average ambient medium there. This asymmetric
ambient medium naturally explains the radio asymmetry.  In several
locations, significant morphological changes and strongly nonradial
motions are apparent. The spatially-integrated X-ray flux continues to
increase with time. Based on {\sl Chandra} observations spanning 8.3
yr, we measure its increase at $1.3\% \pm 0.8\%$ yr$^{-1}$.  The SN
ejecta are likely colliding with the asymmetric circumstellar medium
ejected by the SN progenitor prior to its explosion.

\end{abstract}


\section{Introduction}
\label{intro}

The youngest Galactic supernova remnant \src\ \citep[][]{reynolds08}
has provided important new information on the very early development
of a supernova remnant (SNR) as it interacts with the immediate SN
environment.  Its rapid expansion \citep[$\sim 13,000$ km s$^{-1}$,
  for a presumed distance of 8.5 kpc;][hereafter C11]{carlton11}
allows the study of expansion over relatively short time baselines.
In \citetalias{carlton11}, we compared observations in 2007 (Epoch I)
and 2009
(Epoch II) in X-rays to obtain a mean expansion rate of 0.642\% $\pm$
0.049\% yr$^{-1}$, or a mean expansion age of $156 \pm 11$ yr.
Assuming a mean deceleration rate $m$ ($R \propto t^m$) of about $m =
0.7$, this implies a remnant age of 100 yr.  However, the expansion is
far from uniform.  A much longer (1 Ms) observation in 2011 (Epoch
III) allowed the detection of departures from uniformity in expansion
along the major axis (Fig.~\ref{radio}): the expansion varies
systematically with radius, decreasing from $0.84\% \pm 0.06\%$
yr$^{-1}$ to $0.52\% \pm 0.03\%$ yr$^{-1}$ \citep[][hereafter
  B14]{borkowski14}.  This difference was ascribed either to asymmetry
in the surrounding medium, or in the ejecta themselves, indicating a
significantly anisotropic explosion.  We have now obtained an
additional observation in X-rays of 400 ks (2015) which allows the
further study of nonhomologous expansion in \src.  These observations
should inform ongoing work on modeling G1.9+0.3 \citep[e.g.,][]
{yang16,chakraborti16,tsebrenko15a,tsebrenko15b}.

The high expansion velocities, absence of an obvious pulsar-wind
nebula, and bilateral symmetry of the
X-ray emission \citep[as in SN 1006;][]{winkler14} suggest that
\src\ originated in a Type Ia event.  The detailed dynamics
can then provide clues to the nature of these explosions, at
an age when ejecta still dominate
the dynamics.  Thermal emission from Si, S, and Fe in isolated
regions has allowed a picture to be drawn of highly anisotropic,
overturned material \citep[][]{borkowski13}.  

One mystery, evident from the earliest X-ray observations, was the
difference in morphology between radio and X-ray.  The radio image
(Fig.~\ref{radio}) shows a broad maximum across the northern rim,
rather than bilaterally symmetric rims in the SE and NW.  Since all
the emission is synchrotron, this morphological contrast is puzzling.
SN 1006, for instance, has virtually identical images in radio and
nonthermal X-rays \citep{winkler14}.  The simplest explanation
involves assuming that radio-bright but X-ray faint regions result
from an energy distribution of relativistic electrons that is larger
than that from X-ray bright regions at radio-emitting energies of
order 1 GeV, but which cuts off before reaching X-ray emitting
energies.  One must then search for a mechanism to produce such
differences.

\section{Observations}
\label{obssec}

\subsection{X-ray}

%\floattable
\begin{deluxetable}{lccc}
\tablecolumns{4}
\tablecaption{{\sl Chandra} Observations of G1.9+0.3 in 2015. \label{observationlog}}
\tablehead{
\colhead{Date} & Observation ID & Roll Angle & Effective exposure time \\
& & (deg) & (ks) }

\startdata
May 04--05   & 16947 & 87  & 39 \\
May 05--07   & 17651 & 87  & 111 \\
May 09--10   & 17652 & 87  & 26 \\
May 20       & 16949 & 75  & 9  \\
Jul 14--15   & 16948 & 272 & 40 \\
Jul 15       & 17702 & 272 & 37 \\
Jul 17       & 17699 & 272 & 20 \\
Jul 24       & 17663 & 272 & 57 \\
Jul 25       & 17705 & 272 & 10 \\
Aug 31       & 17700 & 260 & 15 \\
Sep 01--02   & 18354 & 260 & 29 \\ 
\enddata

\end{deluxetable}

The most recent (Epoch IV) {\sl Chandra} observations of
\object{G1.9+0.3} took place in 2015 Spring and Summer in 11
individual pointings (Table \ref{observationlog}) with a combined
effective exposure time of 392 ks, with the remnant again placed on
the Advanced CCD Imaging Spectrometer (ACIS) S3 chip. In order to
reduce particle background, Very Faint Mode was used.  All
observations were reprocessed with CIAO v4.7 and CALDB 4.6.7, then
spatially aligned to the longest pointing from early May (Obs.~ID
17651) using photons from the remnant itself (see
\citetalias{carlton11} and \citetalias{borkowski14} for description of
the alignment method). Numerous point sources present near
\object{G1.9+0.3} were then used to perform the final alignment to the
deep (977 ks) Epoch III observation. The exposure-weighted time
interval $\Delta t$ between Epochs III and IV is 4.057 yr. While
shorter 49.6 ks Epoch I and 237 ks Epoch II observations provide
longer time baselines (8.331 yr and 5.918 yr, respectively), this
advantage is more than offset by the increased counting noise compared
to the long Epoch III observations. Therefore, the expansion
measurements described here are based on the Epoch III and IV
observations alone.

Images, $512^2$ in size, were extracted from the merged event files by
binning event positions to half the ACIS pixel size, so one image
pixel is $0 \farcs 246 \times 0 \farcs 246$. We also extracted data
cubes, $512^2 \times 15$ in size, using the same spatial pixel size,
while spectral channels from 76 to 555 (1--8 keV energy range) were
binned by a factor of 32.

X-ray spectra were extracted from individual rather than merged event
files, and then summed (response files were averaged).  Spectral
analysis was performed with XSPEC v12.8.1 \citep{arnaud96}, using
C-statistics \citep{cash79}. Background was modeled rather than
subtracted.  Spectra of \object{G1.9+0.3} were modeled with an
absorbed power law, using solar abundances of \citet{grsa98} in the
{\tt phabs} absorption model.

\subsection{Radio}

\begin{figure}
  \hspace*{-4mm}
\epsscale{1.26}
%\plotone{f1.eps}
\plotone{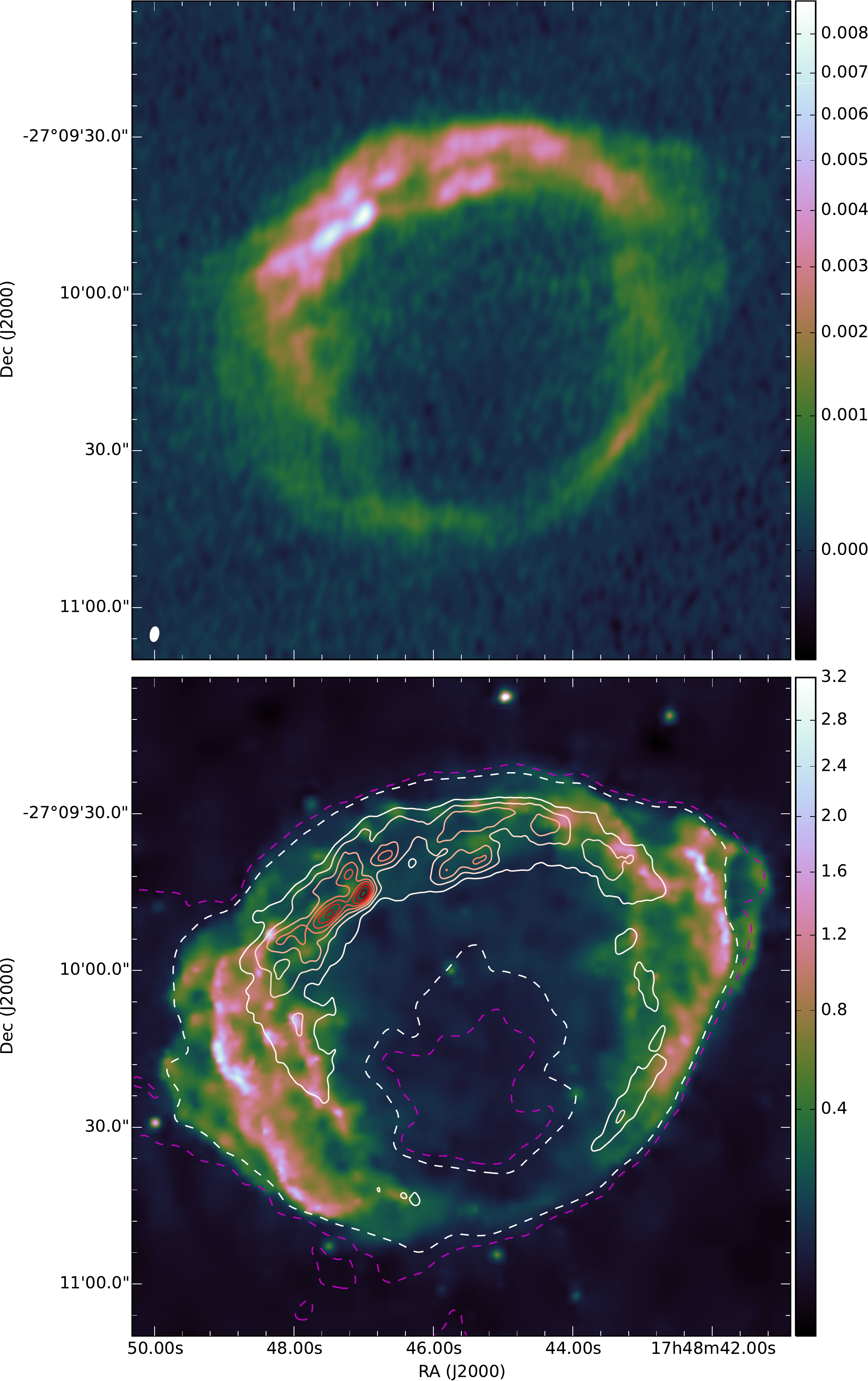}
%\plotone{g1p9vla08.eps}
\caption{Top: Total intensity VLA image of G1.9$+$0.3 at 1365~MHz. The
  resolution is $2 \farcs 8 \times 1 \farcs 6$ at a PA of $-9\fdg6$ (as shown
  in the left-hand corner). The scale is in Jy beam$^{-1}$. Bottom: Smoothed
  2009 1.2 -- 8 keV {\sl Chandra} image overlaid with selected radio contours
  emphasizing bright (solid lines from 1 to 8 mJy beam$^{-1}$ spaced by 1 mJy
  beam$^{-1}$) and very faint (dashed lines in magenta and white at 0.06 and
  0.12 mJy beam$^{-1}$) emission. The scale is in counts per
  $0 \farcs 246 \times 0 \farcs 246$ image pixel (half an ACIS pixel).
  Intensities are shown with the cubehelix color scheme of \citet{green11}.}
\label{radio}
\end{figure}

G1.9$+$0.3 was observed with the {\sl Karl G.~Jansky} Very Large Array
(VLA) in three configurations between 2008 December and 2009 July, in
L-band. Two observations were made in A array, each of 5.7~hr on
source, and one each in the B and C array, each of 0.6~hr on
source. The observations were made with two 25~MHz
bandwidths. However, one of these was badly affected by interference,
so the results presented here are from a single 25~MHz bandwidth,
centered at 1365~MHz. The observations were calibrated using standard
techniques in the Astronomical Image Processing System
(\textsc{AIPS}), and combined for imaging. Figure~\ref{radio} shows
the image of G1.9$+$0.3, with a resolution of $2 \farcs 8 \times 1
\farcs 6$ at a PA of $-9\fdg6$, which has an r.m.s.\ noise of $\approx
0.054$~mJy~beam$^{-1}$. The inclusion of the smaller configurations in
these observations, i.e.\ B and C array, means that this image is
sensitive to structures on all scales from G1.9$+$0.3. Radio contours,
overlaid over a smoothed 2009 {\sl Chandra} image in Figure
\ref{radio}, show that the bright radio emission in the north is
bounded on the outside by faint X-ray emission that marks the location
of the primary blast wave there. Similarly, radio emission in the SE
and NW is also bounded on the outside by X-ray filaments, but this
time X-rays are bright and the radio emission is faint. Radio contours
at low surface brightness (Fig.~\ref{radio}) show that this contrast
between bright X-rays and faint radio is most conspicuous for the
outermost protrusions (``ears'') in the SE and NW.

\section{Expansion} \label{expansion}

Relatively simple, parametric methods for measuring expansion (such as
those used in \citetalias{carlton11} and \citetalias{borkowski14}) cannot
describe the very complex expansion of \object{G1.9+0.3} revealed by
the new {\sl Chandra} observations.  We now use the ``Demons'' algorithm
of \citet{thirion98}, as implemented in the SimpleITK software package
\citep{lowekamp13}, to measure the nonuniform expansion of
\object{G1.9+0.3}.  Attributing morphological changes only to motions,
the Demons method provides a nonparametric way for measuring these
motions globally. Since this computationally-efficient method
(approximately) minimizes the sum of squared differences between pixel
intensities, we need to apply it to the smoothed images, not to raw
{\sl Chandra} images that are badly affected by counting noise.

We smoothed {\sl Chandra} data with the non-local PCA method of
\citet{salmon14}, with smoothed images extracted from smoothed data
cubes.  This method combines elements of dictionary learning and
sparse patch-based representation of images (or spectral data cubes)
for photon-limited data.  Because this Poisson-PCA method is
computationally intensive, relatively small ($512^2 \times 15$) data
cubes, heavily binned along the spectral dimension as described above,
were smoothed using patches $10^2 \times 6$ in size. The
moderate spatial patch size of $2 \farcs 46 \times 2 \farcs 46$
preserves sharp spatial structures seen in the bright filamentary
features, while a large patch size in the spectral dimension is
suitable for the synchrotron-dominated spectra of \object{G1.9+0.3} that
vary smoothly across the entire spectral range of {\sl
  Chandra}. With the patch size chosen, the most important parameters
that control the smoothing of data cubes are the rank $l$ of the
approximation to the underlying intensity to the collection of patches
(order of the Poisson-PCA method), and the number of clusters $K$ into
which patches are grouped prior to estimation of intensities. We used
$l=6$ and $K=30$ for the 2011 and 2015 data cubes of
\object{G1.9+0.3}.

\begin{figure}
  \hspace*{-4mm}
\epsscale{1.26}
%\plotone{f2.eps}
\plotone{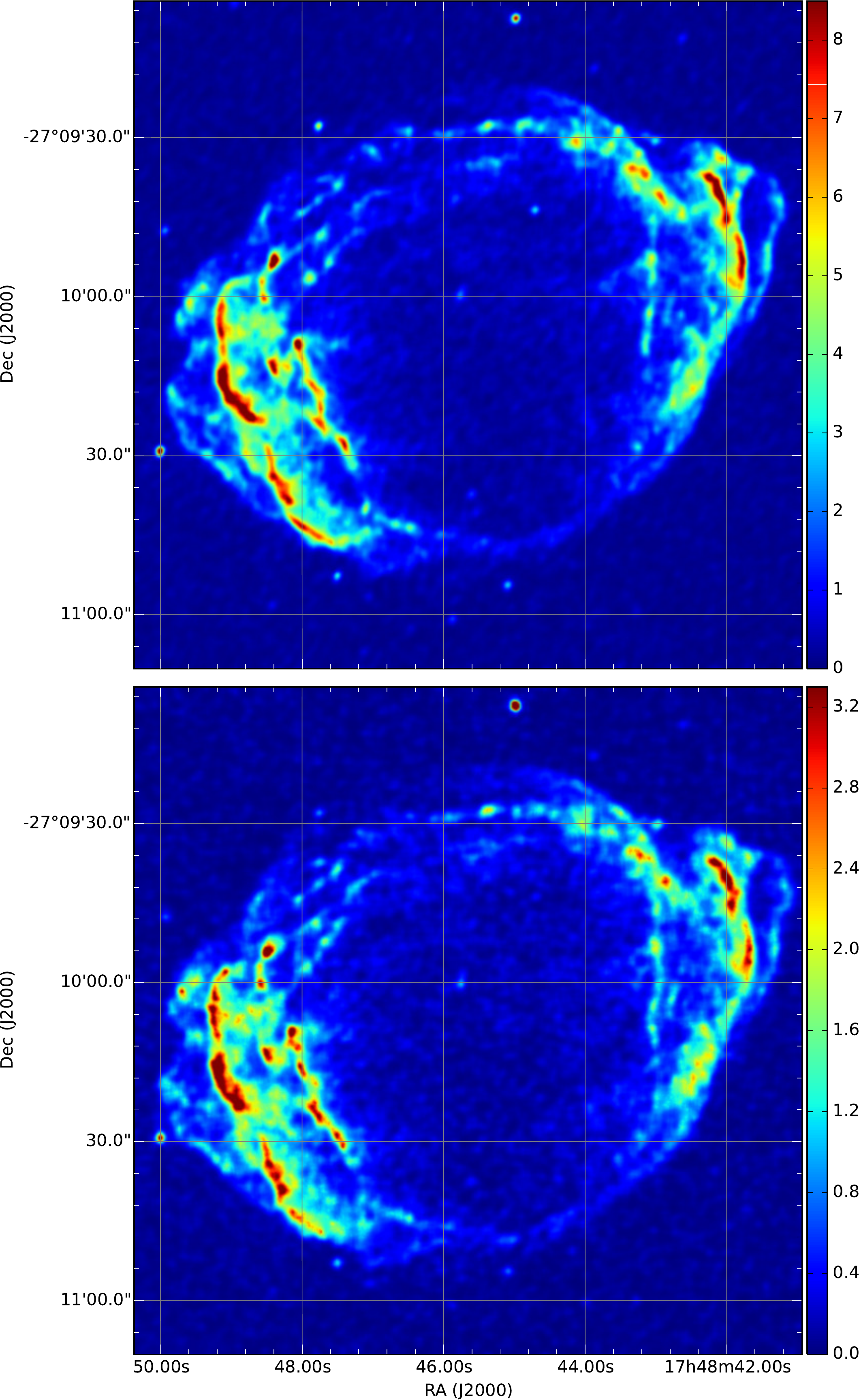}
%\plotone{xrays11and15.eps}
\caption{Smoothed X-ray images of G1.9+0.3 in the 1--8 keV energy range from
2011 (top) and 2015 (bottom), with a coordinate grid
superposed. Expansion between these Epoch III and IV observations is
apparent. The scale is in counts per $0 \farcs 246 \times 0 \farcs 246$
image pixel.}
\label{xrays11and15}
\end{figure}

\begin{figure}
\epsscale{1.22}
%\plotone{f3.eps}
\plotone{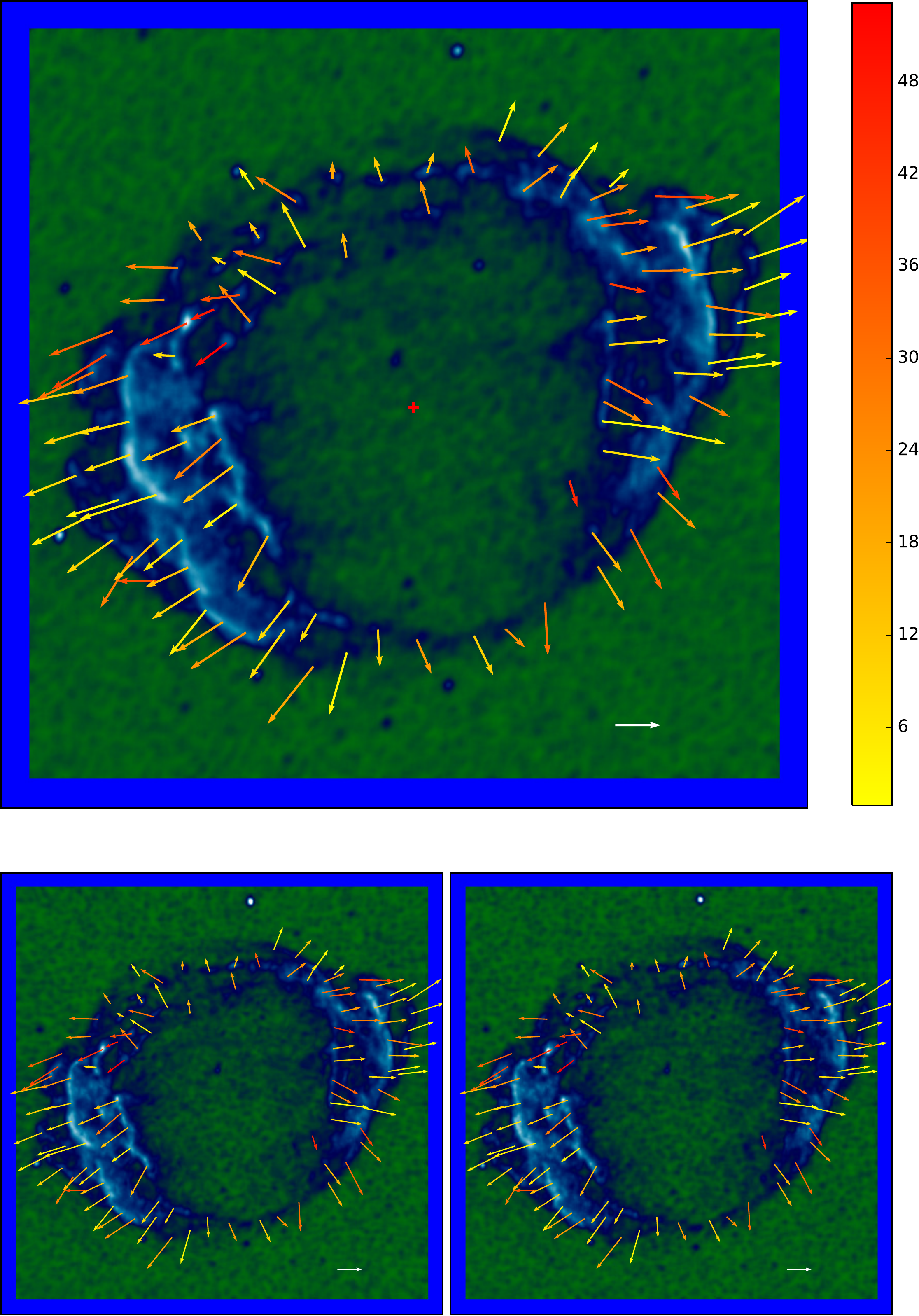}
%\plotone{g1p9propermotionsangles.eps}
\caption{Proper motion vectors overlaid over the 2011 (top), 2015 (bottom
  right), and the transformed 2015 (bottom left) images. They are
  color coded according to
  the deviations in direction from radial (with respect to the geometrical
  center of the remnant, marked by the red cross), in degrees according to the
  vertical scale. These proper motion data are available as the Data behind the
  Figure. The white arrows indicate $0 \farcs 25$ yr$^{-1}$. }
\label{propermotions}
\end{figure}

The expansion of \object{G1.9+0.3} is readily apparent by comparing
smoothed {\sl Chandra} images from 2011 and 2015
(Fig.~\ref{xrays11and15}). We measure an average expansion rate of
0.58\%\ yr$^{-1}$ between these 2 epochs
\citepalias[using the same method as in][]{carlton11},
but there are very large deviations from uniform expansion.
In only 4 years, striking morphological variations at various locations
within the remnant can be discerned by eye. These include bending of
the innermost rim in the west, and even more complex, strongly
nonradial motions in the NE where changes in relative positions of
prominent emission knots and filaments can be seen.  In the framework
of the Demons algorithm, which can account for such complex
motions, the 2011 1--8 keV smoothed image is considered as the {\sl
reference} image, while the 2015 1--8 keV smoothed image is the {\sl
moving} image. Before application of the Demons method, particle
background with an estimated rate of $9.0 \times 10^{-5}$  
and $7.0 \times 10^{-5}$ cts s$^{-1}$
per image pixel was subtracted from the 2011 and 2015 images respectively,
and then each background-subtracted image was normalized using a
corresponding monochromatic ($E = 3$ keV) exposure map. Point sources
and the featureless interior of the remnant have been masked out.
The uniform expansion at
0.58\%\ yr$^{-1}$ centered $10 \farcs 5$ north of the geometrical center
was used to set initial displacements between the two epochs, then
at each iteration step of the Demons method computed displacements
were smoothed with a Gaussian with FWHM of
$0 \farcs 82$\footnote{Other parameters of the Demons method
are far less important and primarily affect its convergence
rate.}.
After achieving convergence, this smoothing resulted in a
good match between the 2011 image and the (transformed) 2015 image
(Fig.~\ref{propermotions}), demonstrating the effectiveness of the Demons
method,  
but the final displacement vectors were affected by substantial
noise. We reduced this noise by calculating brightness-weighted
averages of displacement vectors within various regions of
interest. Finally, these averaged displacements were divided by
$\Delta t$ to obtain proper motion vectors $\vec{\mu}$.

We show vectors $\vec{\mu}$ in Figure~\ref{propermotions} for 
a large number of regions chosen to delineate all major morphological
features of \object{G1.9+0.3}.
The white arrow with length $0 \farcs 25$
yr$^{-1}$ (10,100 km s$^{-1}$ at 8.5 kpc) shows their scale, while their tails
are located at the geometrical centroids of these regions. Motions are
between $0
\farcs 09$ yr$^{-1}$ and $0 \farcs 44$ yr$^{-1}$, varying by a factor
of up to 5. The median $\mu$ is $0 \farcs 28$ yr$^{-1}$.  The
slowest shocks are in the north.  The fastest motions are
predominantly along the major axis of the remnant.  Motions are
strongly nonradial with respect to the remnant's geometrical center
at R.A.~$17^h48^m45 \fs 61$, Decl.~$-27 \arcdeg 10
\arcmin 05 \farcs 6$.
We infer the explosion occurred somewhere NE of the
geometrical center.  Extreme deviations from radial motions are
present.

\begin{figure}
  \hspace*{-7mm}
\epsscale{1.28}
%\plotone{f4.eps}
\plotone{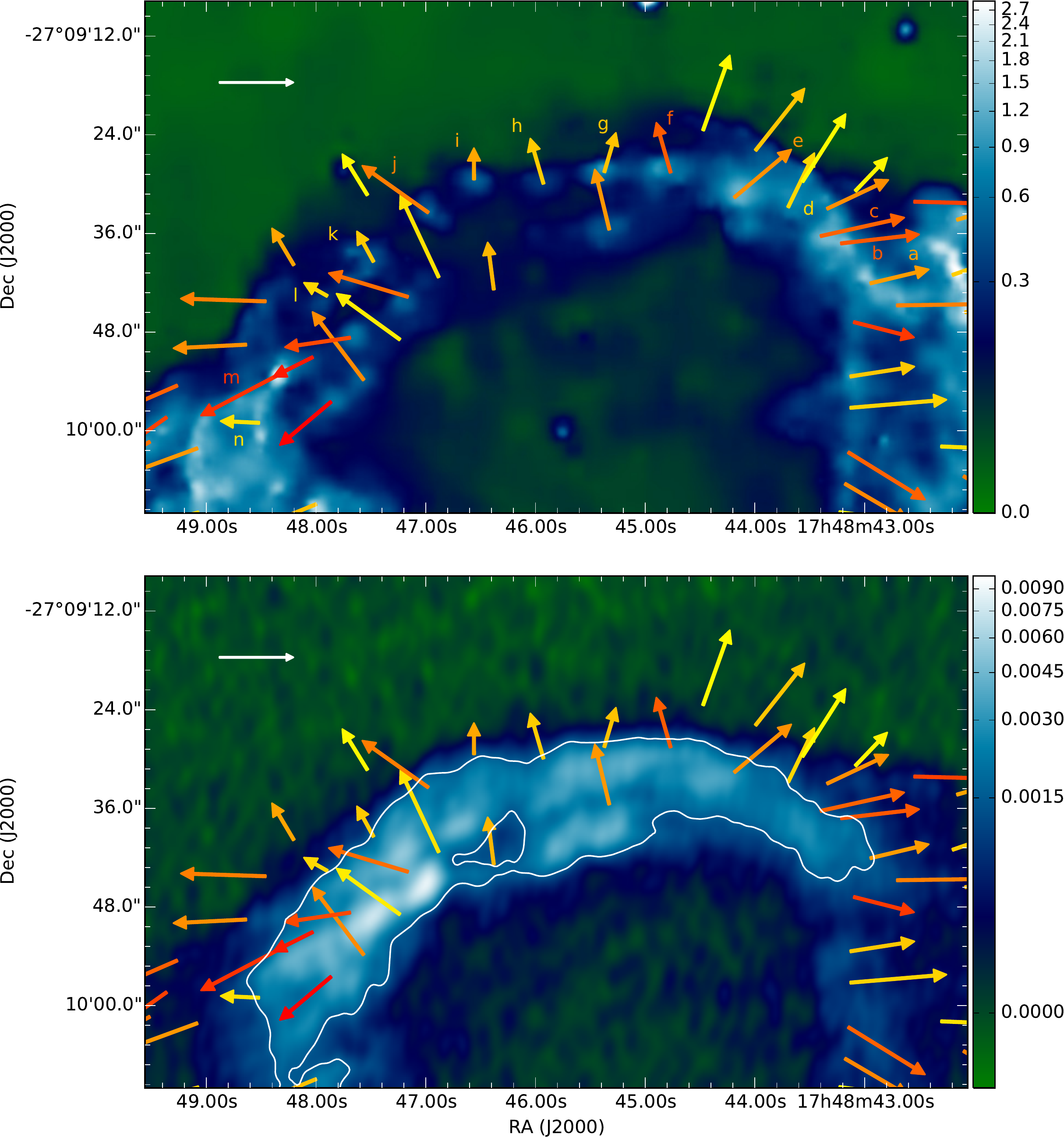}
%\plotone{xrays09vla08.eps}
\caption{2009 X-ray (top) and radio (bottom) images centered on the
  radio-bright northern rim, with X-ray proper-motion
  vectors overlaid (color coded as in Fig.~\ref{propermotions}). The white
  arrows indicate $0 \farcs 25$ yr$^{-1}$. The radio contour is at 1.5 mJy
  beam$^{-1}$. Selected vectors along this contour have been labeled.
  The scales are in the same units as
  in Figure~\ref{radio}. Note the very low velocities just ahead of the
  outermost bright radio emission. }
\label{r-xmotions}
\end{figure}

In Figure~\ref{r-xmotions}, we draw vectors $\vec{\mu}$ over
the radio and contemporaneous 2009 {\sl Chandra} images (tails of
vectors were shifted inwards to account for the remnant's expansion
between the 2008 December array A VLA and Epoch III {\sl Chandra}
observations). Expansion is noticeably slower along the outer boundary
of the radio emission. The median
motion along the outer radio
contour at 1.5 mJy beam$^{-1}$ (regions {\sl a} to {\sl n} in
Fig.~\ref{r-xmotions}) is $0 \farcs 17$ yr$^{-1}$ (only 6,900 km
s$^{-1}$), significantly less than the overall median of $0 \farcs 28$
yr$^{-1}$ (11,000 km s$^{-1}$). Among all regions with
$\mu < 0 \farcs 17$ yr$^{-1}$, most (12 out of 14) are within the
northern rim, predominantly toward the N and NE where radio emission
is particularly bright.  This includes regions {\sl f} ($0 \farcs 16$
yr$^{-1}$), {\sl g} ($0 \farcs 13$ yr$^{-1}$), {\sl h} ($0 \farcs 14$
yr$^{-1}$), {\sl i} ($0 \farcs 095$ yr$^{-1}$), {\sl k} ($0 \farcs 11$
yr$^{-1}$), {\sl l} ($0 \farcs 089$ yr$^{-1}$), and {\sl n} ($0 \farcs
13$ yr$^{-1}$). The smallest $\mu$ (region {\sl l})
corresponds to a velocity of only 3,600 km s$^{-1}$. But motions of
several knots and filaments in the NE are larger, more typical for the
remnant as a whole, including regions {\sl j} ($0 \farcs 26$
yr$^{-1}$) and {\sl m} ($0 \farcs 29$ yr$^{-1}$). The latter
corresponds to a very bright X-ray knot with motion strongly
deviating from radial (by $42^\circ$), while the former is the
northernmost of 5 fast-moving knots/filaments that are mostly located
near where G1.9+0.3 is the brightest in radio.

The motions along the NW outer radio contour (regions {\sl a -- e}) are
$0 \farcs 20$ yr$^{-1}$, $0 \farcs 27$ yr$^{-1}$, $0 \farcs 29$ yr$^{-1}$,
$0 \farcs 19$ yr$^{-1}$, and $0 \farcs 24$ yr$^{-1}$, respectively, with
a median of $0 \farcs 24$ yr$^{-1}$ (9,700 km s$^{-1}$). This is significantly
more than in the N -- NE, where the median for regions {\sl f -- n} is
$0 \farcs 13$ yr$^{-1}$ (5,300 km s$^{-1}$), but still less than average for 
G1.9+0.3. As in the NE, strongly nonradial motions are present. Unlike
the N -- NE where X-rays are generally faint and radio is quite bright, this
region is moderately bright both in X-rays and radio.

The motions along the major axis are faster than the overall median of
$0 \farcs 28$ yr$^{-1}$. For the brightest
X-ray rims in the middle, the median is $0 \farcs 32$ yr$^{-1}$, 
increasing modestly to $0 \farcs 33$ yr$^{-1}$ ahead of these rims. For the
innermost rims \citepalias[this includes vectors $\vec{\mu}$ within the inner 
pair of regions in][]{borkowski14}, the median drops to
$0 \farcs 30$ yr$^{-1}$. These rather modest (12,000 -- 13,000 km s$^{-1}$)
radial variations along the major axis are not surprising in view
of the rapid radial decrease in expansion rates found by
\citetalias{borkowski14}.

For the southern radio rim, the
motion is $0 \farcs 21$
yr$^{-1}$ (the median for 7 regions along this rim). This is
significantly more than $0 \farcs 13$ yr$^{-1}$ that we found in the N
-- NE, and corresponds to velocity of 8,300 km s$^{-1}$. The
southernmost extension toward the east, with only a very faint radio
counterpart, must be moving much faster than this as we
find $\mu = 0 \farcs 38$ yr$^{-1}$ (15,000 km
s$^{-1}$). Apparently, the motions are also complex in this relatively
faint region of the remnant.

\section{Flux Increase} \label{fluxincrease}

A joint fit to the spatially-integrated 1--9 keV spectrum of
\object{G1.9+0.3} and to the background spectrum gave a best-fit
photon index $\Gamma$ of $2.45 \pm 0.06$ and an absorbing column
density $N_H = 7.41 (\pm 0.16) \times 10^{22}$ cm$^{-2}$, with an
absorbed 1--7 keV flux $F_{\rm 2015}$ of $3.063 (\pm 0.027) \times
10^{-12}$ erg cm$^{-2}$ s$^{-1}$ (errors are 90\%\ confidence
intervals). Both $\Gamma$ and $N_H$ are consistent with previous
measurements \citepalias[$\Gamma = 2.40 \pm 0.03$ and $N_H = 7.25 (\pm 0.09)
\times 10^{22}$ cm$^{-2}$;][]{borkowski14}, while the flux increased by
3\%\ since 2011. So \object{G1.9+0.3} continues to increase in
brightness, but at a somewhat lower rate than the $1.9\%$ yr$^{-1}$
reported in \citetalias{borkowski14}.

A linear regression fit to $F_{\rm 2015}$ and to the
previously-measured fluxes \citepalias[$F_{\rm 2007}=2.73 \times
10^{-12}$ erg cm$^{-2}$ s$^{-1}$, $F_{\rm 2009}=2.88 \times 10^{-12}$
erg cm$^{-2}$ s$^{-1}$, $F_{\rm 2011}=2.97 \times 10^{-12}$ erg
cm$^{-2}$ s$^{-1}$;][]{borkowski14} gives a rate of $1.3\% \pm 0.8\%$ yr$^{-1}$.  Flux
residuals (model fit minus observations) range from $-1.4\%$ (2011) to
$1.2\%$ (2007), comparable to the ($1\sigma$) statistical error of
$1.3\%$ for $F_{\rm 2007}$ but more than $0.3\%-0.6\%$ found at other
epochs. Although only about 8000 source counts were detected in the
short 2007 observation, systematic errors are already becoming
comparable to statistical errors for {\sl Chandra} spectra with that
number of counts ($\sim 10^4$).
The much longer observations from later epochs are then dominated by
systematic (not statistical) errors, and the errors are likely
comparable for all 4 epochs.  In this case, the flux residuals quoted
above are consistent with a linear flux increase from 2007 to
2015. Our newly-determined, more accurate rate of $1.3\%$ yr$^{-1}$,
with the relatively large error of $\pm 0.8\%$ yr$^{-1}$ accounting
for systematic effects, is also consistent with the larger rate of
$1.9\%$ yr$^{-1}$ ($\pm 0.4\%$ yr$^{-1}$ -- statistical errors only)
based on the Epoch I--III observations alone \citepalias{borkowski14}.

\section{Discussion}

Figure~\ref{r-xmotions} shows that the slowest expansion speeds are
found just beyond the radio maxima.  Evidently the shock is
encountering denser material there, resulting in a drastic
deceleration of the expansion.  Such denser material in a slower shock
will have two effects on the accelerated electron spectrum: a larger
number density of accelerated electrons \citep[since that density is
  generally found to be a fraction $\sim 10^{-4}$ of the total
  particle density; e.g.][]{ellison05}, and a lower maximum energy,
since $E_{\rm max} \propto B v_{\rm shock}^2$ for age-limited
acceleration \citep[e.g.,][]{reynolds08a}.  The resulting synchrotron
emission will be brighter in radio than that from
surrounding regions with faster shocks, but the spectrum will turn
over at lower energies and may not reach to X-rays.  That is, the
slower shocks we observe, produced by denser ambient material, can
explain the differing radio and X-ray morphologies.  Assume
(conservatively) $B \propto \rho^{1/2}$ (no extra
magnetic-field amplification), then since 
we expect $\rho v_{\rm shock}^2 \sim $ const., we have $E_{\rm max}$
(age) $ \propto \rho^{-1/2} \propto v_{\rm shock}$, and $\nu_{\rm
  roll} \propto E_{\rm max}^2 B \propto v_{\rm shock}$.  In the region
of the radio maximum, the mean shock speed is about 4,000 km s$^{-1}$,
or about 0.3 of the value at the bright rims.  The integrated spectrum
of G1.9+0.3 is well described by a rolloff of $3.1 \times 10^{17}$ Hz
\citep{zoglauer15}, implying a rolloff near the radio peak of about
$9 \times 10^{16}$ Hz or less than 0.4 keV -- too low to produce
appreciable X-ray synchrotron emission at that location.

The rate of flux increase we find, 1.3\% yr$^{-1}$, is still in the
range of radio increase rates of 1 -- 2\% yr$^{-1}$
\citep{green08,murphy08}, so the important question of the comparison
of radio and X-ray rates is still unsettled.  Different rates imply
time-variation in the maximum electron energies, or something even
more unexpected.  Further observations of both radio and X-rays will
be crucial.

The large range of velocity vectors, in both magnitude and direction,
that we find in G1.9+0.3 highlights the inadequacy of spherically
symmetric models in describing SN events and their aftermath.  The
north rim is particularly poorly described by 1-D models.  Failure to
account for the large variations in velocity casts doubt on such
analyses.  However, these velocities contain a great deal of
information on the explosion and the circumstellar environment that
rivals Kepler's SN, the most recent historical SN in the Galaxy, in
its complexity.  As in Kepler's Type Ia SNR, large ($> 10$) density
gradients are present, and SN ejecta are likely colliding with the
asymmetric circumstellar medium ejected by the SN progenitor.  Our
statistical method of determining velocities throughout an expanding
remnant, and not just in well-defined sharp outer edges, should have
wide application in following the expansion of young SNRs and
elucidating their early evolution and the nature of the surrounding
medium.

\acknowledgments
We acknowledge support by NASA through {\sl Chandra} General Observer
Program grants SAO GO5-16069A -- C.

\vspace{5mm}
\facilities{CXO, VLA}

%\bibliography{g1p9}
%\end{document}

\end{document}